\documentclass{aa}  
\usepackage{graphicx}
\usepackage{natbib}
\usepackage{txfonts}
\begin{document}
   \title{Constituents of the soft X-ray background}

   \author{A. M. So\l tan         \inst{}                }

   \offprints{A. M. So\l tan}

   \institute{Nicolaus Copernicus Astronomical Center,
              Bartycka 18, 00-716 Warsaw, Poland\\
              \email{soltan@camk.edu.pl}                 }

   \date{Received \hspace{15mm}   ; accepted}

  \abstract
   {The X-ray background is generated by various classes of objects
    and variety of emission mechanisms. Relative contribution of
    individual components depends on energy. }
   {The goal is to assess the integral emission of the major components of
    the soft X-ray background (extragalactic discrete
    sources dominated by AGNs, galactic plasma, and the Warm/Hot Intergalactic
    Medium), investigating the angular structure of the background.}
   {Fluctuations of the background are measured using the auto-correlation
    function of the XRB determined in 5 energy bands between
    $0.3$ and $4.5$\,keV. The investigation is based on the extensive observational
    data set selected from the XMM-Newton archives. }
   {Amplitudes of the auto-correlation functions calculated in three
    energy bands above $\sim\!1$\,keV are consistent with the conjecture
    that the background fluctuations result solely from clustering of
    sources which produce the background. At energies below $1$\,keV
    the relative fluctuation amplitude decreases indicating that
    a fraction of the soft XRB is associated with a smooth plasma
    emission in the Galaxy. It is shown, however, that the mean spectrum
    of extragalactic discrete sources steepens in the soft X-rays and
    is not well represented by
    a single power law in the energy range $0.3-4.5$\,keV. The
    WHIM contribution to the total background fluctuations is small
    and consistent with the WHIM properties derived from the cross-correlation
    of the XRB with galaxies. }
   {}


   \maketitle
%

\section{Introduction}

The background in the 'classic' X-ray domain of $2-8$\,keV is mostly generated
by discrete extragalactic sources (e.g. \citealt{lehmann01, kim07}, and
references therein).  At lower energies, particularly below $1$\,keV, diffuse
hot plasma emission in the Galaxy and in the intergalactic space make a
detectable contribution (e.g.  \citealt{henley07}). The objective of the paper
is to assess amplitudes of major X-ray background (XRB) components in different
energy bands. The method exploits fluctuation analysis and dependence of the
correlation functions of the XRB on the energy band. No assumptions are made on
the spectral shape of the individual components.

Time-consuming extensive observations of a large volume of individual objects
and precise measurements of the XRB spectrum are considered the
straightforward way to determine absolute fluxes of various components
contributing to the XRB. However, this method involves detailed modeling of
physical parameters of the emitting gas and assumptions on the extragalactic
component (cf \citealt{mccammon02,henley07}). Here we resign from estimates of
the physical status of the galactic and extragalactic media and concentrate on
the overall amplitudes of major constituents of the XRB. 

The total XRB flux is a sum of several components which have different spatial
distributions. Consequently, the resultant XRB fluctuations depend on the
relative strength and distribution of each component.  In this investigation we
distinguish three basic constituents of the XRB.  First, the extragalactic
point-like sources, which are known to dominate the XRB at higher energies.
Fluctuations generated by these objects at angular scales larger than the width
of the Point Spread Function (PSF) are described by the relevant
auto-correlation function. Second, the galactic emission generated by the Local
Bubble and the halo (e.g. \citealt{galeazzi07}). Although the surface
brightness distribution of the Local Bubble is highly anisotropic at large
angular scale, presumably it is smooth at small scales. It is expected that the
halo emission is also smoothly distributed.  Third, the emission produced by
the intergalactic medium.

A major fraction of baryons in the local universe remains in the intergalactic
space (e.g.  \citealt{cen99}). In the recent years, physical conditions of this
diffuse matter component have been investigated theoretically and examined
observationally by several groups. Simulations of interactions between
primordial gas falling onto galaxies with galactic wind indicate that some
fraction of the diffuse component is heated in shocks and enriched with heavier
elements (e.g.  \citealt{dave01,bryan01,croft01}). In the areas around
concentrations of matter, density of the diffuse component increases
substantially and temperatures reach $10^5 - 10^7$\,K.  Because of
characteristic physical parameters, this component of baryonic matter has been
termed by \cite{dave01} as {\em Warm/Hot Intergalactic Medium} (WHIM).  At
temperatures above $\sim\!10^6$\,K the WHIM becomes a source of soft X-rays
emitted via the thermal Bremsstrahlung mechanism. Thus, the WHIM contributes to
the X-ray background (XRB) apart from well recognized sources. However, the
WHIM signal even at low energies is expected to be small in comparison to the
integral XRB surface brightness. It is completely negligible above $2$\,keV. At
lower energies the WHIM contribution increases, but using only the spectral
characteristics, it is difficult to distinguish this component from the
emission by hot plasma in the Galaxy halo.

Spatial distribution of the WHIM is highly nonuniform. Clumpiness of the WHIM
emission is a crucial factor which allowed us to isolate and identify this
constituent of the XRB and observationally confirm existence of the WHIM itself
(\citealt{soltan02,soltan05}). Both the soft spectrum and large linear sizes of
the emitting regions are consistent with the theoretical predictions for the
WHIM clouds based on the hydrodynamical simulations. Observations also confirm
theoretical estimates that the WHIM contribution to the integral XRB flux is
low even in the soft energy bands (see below). Consequently, the WHIM
contribution to the fluctuation amplitude of the integral XRB is also
small in comparison to the fluctuations generated by AGNs.

In the present paper the auto-correlation function (ACF) of the total counts,
including the non-cosmic contamination is determined in five energy bands
covering range $0.3 - 4.5$\,keV.  A relationship between the ACF signal of the
raw counts and the intrinsic ACF amplitude of the XRB is obtained. To evaluate
a cosmic signal, the ACF requires `suitable' data, corrected for observational
and instrumental bias. A question of what sources generate the XRB
fluctuations is discussed.

The organization of the paper is following.  In the next section procedures and
criteria applied to the observational material are presented. In
Sect.~\ref{acf} a method to calculate the ACF using a large set of pointing
observations is described and the raw results of the calculations are
presented. The interpretation of the ACF amplitude  distribution over the total
investigated energy range is examined in Sect.~\ref{models}. Basic conclusions
of the paper are summarized in Sect.~\ref{conclusions}.

\section{Observational material}

The X-ray data are extracted from the public archive\footnote{XMM-Newton
Science Archive:\\
http://xmm.vilspa.esa.es/external/xmm\_data\_acc/xsa/index.shtml} of the
XMM-Newton EPIC/MOS observations.The pointings are selected in the
similar way as in the previous papers in this series. For the full
description of the data selection and reduction the reader is referred
to papers by \cite{soltan05} and \cite{soltan06}. Here only the main
points are recalled and important modifications of the original
procedures described.

All the data obtained with the MOS1 and MOS2 detectors and the thin
filter have been inspected, and only pointings without strong sources
and known extended sources have been accepted for further processing.
Five energy bands as in \cite{soltan06} have been used:  $0.3-0.5$\,keV,
$0.7-1.0$\,keV, $1.0-1.35$\,keV, $1.9-3.0$\,keV, and $3.0-4.5$\,keV.
Energy gaps $0.5-0.7$\,keV and  $1.35-1.9$\,keV reduce confusing Galaxy
contribution and strong internal background due to fluorescent lines
(\citealt{nevalainen05}). The count distribution for each pointing and
the energy band has been corrected for vignetting (\citealt{soltan05}).

Homogeneity of the data constitutes an obvious criterion which should be
satisfied in the ACF calculations. Criteria applied by \cite{soltan06} to
select the data for the investigation of the cross-correlation function (CCF)
of the XRB and the galaxy distribution are not sufficiently stringent for the
ACF calculations. This is because the ACF is highly vulnerable to various
instrumental effects which are of the lesser importance in the CCF
calculations. In particular, different amount of the particle background in
individual pointings and deficiency in the vignetting corrections have little
effect on the CCF amplitude, but generate large spurious ACF signal. The first
point is dealt with in the next section. To minimize effects of residual
fluctuations due to inaccuracies of the vignetting corrections, we limit the
ACF calculations to the central CCD detector in the both EPIC MOS cameras. In
the central region of the field of view vignetting effects are relatively
small, and the procedure which corrects the vignetting is more reliable.
Moreover, one can expect that the data collected within a single detector are
more homogeneous that those from several detectors. Consequently, the data
restricted to the central detector are less affected by the instrumental biases
and more `suitable' for the fluctuation analysis than all the data recorded in
the `Full Frame' mode. Unfortunately, volume of the usable data is much smaller
and the results of the present calculations are subject to larger statistical
uncertainties.

\section{Auto-correlation function from the scattered pointings \label{acf}}

Constraints imposed on the present observational material severely affect
methods of the ACF calculations. First, each pointing observation covers a
small area of the sky, roughly $10\arcmin \times 10\arcmin$. Second, each
pointing is contaminated by the unknown a priori number of the non-cosmic
counts (mostly charged particles). It implies that the recorded count rates in
two observations cannot be used simultaneously to assess the ACF, or --
equivalently -- the ACF cannot be determined at separations greater than the
angular size of the individual pointing.

Let $t_i(\vec{x})$ represents the distribution of count rates on the celestial
sphere in the $i$-th observation, where $\vec{x}$ defines a point on the sphere.
The total count rate is a sum of the X-ray counts $\rho_i$ and the contaminating
background counts $b_i$:

\begin{equation}
\label{def}
t_i(\vec{x}) = \rho_i(\vec{x}) + b_i\,,
\end{equation}
where both the XRB counts and the total counts depend on the position
$\vec{x}$, while the non--X-ray counts $b_i$ are randomly distributed in the
field of view and are different for each pointing. To find estimate of the XRB ACF
using set of $N$ pointings ($i = 1, ..., N$) we define a function $\Delta_i(r)$:

\begin{equation}
\label{delta}
\Delta_i(r) \equiv \langle t_i(\vec{x})\cdot t_i(\vec{x}+\vec{r})\rangle - 
              \langle t_i \rangle ^2 =
              \langle \rho_i(\vec{x})\cdot \rho_i(\vec{x}+\vec{r})\rangle -
              \langle \rho_i \rangle ^2\,,
\end{equation}
where the angle brackets $\langle ... \rangle$ denote the averaging over the
field of view of the $i$-th pointing. The right-hand side of the Eq.~\ref{delta}
is now expressed as:

\begin{equation}
\label{delta_1}
\Delta_i(r) = \frac{1}{A_i} \int\limits_{A_i} d^2\!x\,\rho_i(\vec{x}) \cdot
              \rho_i(\vec{x} + \vec{r}) - \left[ \frac{1}{A_i}
              \int\limits_{A_i} d^2\!x\,\rho_i(\vec{x})\right]^2\,,
\end{equation}
where $A_i$ is the area of the $i$-th pointing. We now calculate the average
of the relevant quantities in Eqs.~\ref{delta} and \ref{delta_1} over all the
pointings. Eq.~\ref{delta_1} gives:

\begin{eqnarray}
\label{ksi}
\Delta(r)\equiv \frac{1}{N} \sum_i \Delta_i(r) & = & 
           \frac{1}{A} \int\limits_A d^2\!x\,
 \langle \rho_i(\vec{x}) \cdot \rho_i(\vec{x}+\vec{r})\rangle -
                                                          \nonumber \\
   &   & \frac{1}{A^2}
   \int\limits_{\;\;\;A}\!\!\int d^2\!x_1\,d^2\!x_2\,
            \langle \rho_i(\vec{x_1)} \cdot \rho_i(\vec{x_2})\rangle \,.
\end{eqnarray}
Here the angle brackets $\langle ... \rangle$ denote the expectation of
the relevant distributions.
The average over $N$ pointings in the right-hand side of the Eq.~\ref{ksi}
has been replaced by the overall expectation values integrated
over the field of view of a single pointing (all the pointings are of the same
shape). Eq.~\ref{ksi} can be rewritten in the form:

\begin{equation}
\label{delta_2}
\Delta(r) = \langle \rho \rangle^2 \, [ \xi(r) - I(A) ]\,,
\end{equation}
where $\langle \rho \rangle$ is the average count rate of the XRB, $\xi(r)$
is the ACF defined as:
\begin{equation}
\label{ksi_2}
\xi(r) = \frac{\langle \rho(\vec{x}) \cdot \rho(\vec{x}+\vec{r})\rangle}
              {\langle \rho \rangle^2} - 1\,,
\end{equation}
and $I(A)$ is the average of the $\xi(r)$ over the pointing field of view $A$:
\begin{equation}
I(A) = \frac{1}{A^2} \int\limits_A d^2 x_1 \int\limits_A d^2 x_2\, \xi(r = 
                 \left|\vec{x_1} - \vec{x_2}\right|)\,.
\end{equation}

The left-hand side of Eq.~\ref{delta} averaged over all the pointings is used
to define the {\em raw correlation function} $w(r)$:
\begin{equation}
\label{w}
w(r) = \frac{\frac{1}{N} \sum\limits_i 
      \langle t_i(\vec{x}) \cdot t_i(\vec{x}+\vec{r})\rangle -\langle t_i \rangle^2}
      {\langle t \rangle^2}\,,
\end{equation}
where $\langle t_i \rangle$ is the average total count rate in the $i$-th
pointing, and $\langle t \rangle$ is the average count rate over all the pointings.
Eqs.~\ref{delta} and \ref{delta_2} give:
\begin{equation}
\label{ksi_3}
\xi(r) = \frac{\langle t \rangle ^2}{\langle \rho \rangle ^2}\, w(r) +I(A)\,.
\end{equation}
Thus, the ACF is undetermined to an arbitrary additive constant $I(A)$. This
is because
variations of count rates only within a single pointing should be attributed to
the actual XRB fluctuations, while differences of the count rates between
pointings are partially generated by the non-cosmic signal. To eliminate
the unknown parameter $I(A)$, incremental form of Eq.~\ref{ksi_3} will be used:
\begin{equation}
\label{ksi_4}
\xi(r_1) - \xi(r_2) = \frac{\langle t \rangle ^2}{\langle \rho \rangle ^2}\, 
            [ w(r_1) - w(r_2) ]\,,
\end{equation}
where both $r_1$ and $r_2$ are within the field $A$. The $w(r)$ function is
presented in the next Subsection and implications for the XRB components imposed by
the ACF calculated using the Eq.~\ref{ksi_4} are discussed in Sect.~\ref{models}.

\subsection{The raw correlation function}

   \begin{figure*}
   \centering
   \includegraphics[width=17.0cm]{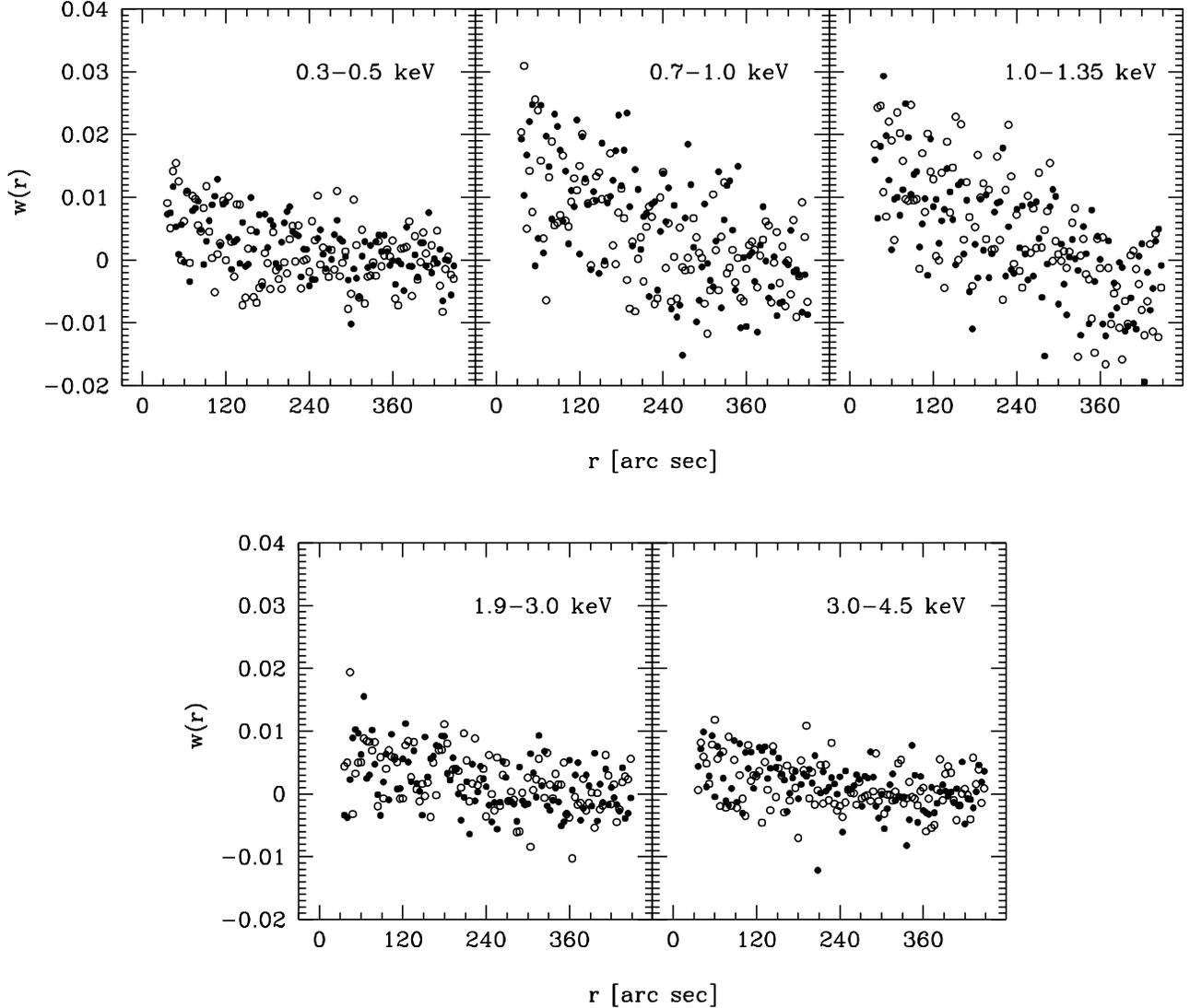}
   \caption{The raw ACF as defined in Eq.~\ref{w}. Open symbols - MOS 1,
filled - MOS 2.}
              \label{raw_acf}%
    \end{figure*}

Nearly 500 pointings have been examined in respect of the use for the present
investigation. About 200 pointings, most of them with two MOS cameras, have
been qualified for further processing. All the point-like sources in the
accepted pointings have been removed.  The observations have been selected at
high galactic latitudes, typically above $\left| b \right| > 30\deg$, and only
the pointings with relatively low hydrogen column density, $N_H$, have been
used. Threshold values for the $N_H$ adopted in the selection process are given
in Table~\ref{delta_w}. Effects related to the low energy absorption in the
Galaxy, albeit small, have been accounted for in the subsequent analysis. The
total exposure time of the data used in the calculations exceeded $8$\,Ms.The
distributions of counts were binned into $4\arcsec \times 4\arcsec$ pixels and
the $w(r)$ functions have been determined according to Eq.~\ref{w}.
Fig.~\ref{raw_acf} shows results of the computations for separations $30\arcsec
< r < 7.5\arcmin$ and 5 energy bands. The $w(r)$ functions  based on MOS 1 and
2 detectors are marked with different symbols. Scatter of points indicates
statistical uncertainties due to the limited number of counts, and both
detectors give similar results. Possible small deviations of the $w(r)$
distributions between MOS 1 and 2 provide some insight into systematic
uncertainties involved in the calculations. All the $w(r)$ distributions
exhibit sharp peak at separations below $\sim\!30\arcsec$ (not shown in plots),
generated by the unresolved sources weaker than the detection threshold. Due to
the wide PSF wings these sources generate a residual signal also above $r =
30\arcsec$. It has been removed in the same way as described by
\cite{soltan06}. Consequently, the amplitude variations of the $w(r)$ functions
shown in Fig.~\ref{raw_acf} reflect the actual large scale fluctuations of the
XRB.

\begin{table}
\caption{Raw correlation function increments}
\label{delta_w}
\centering
\begin{tabular}{clcc}
\hline\hline
\noalign{\smallskip}
 $\#$ &     $E$ ~[keV] & ${N_H}_{\rm max}$~[$10^{21}$cm$^{-2}$] & $\Delta w$ \\
\hline
\noalign{\smallskip}
   1  &  $0.3 - 0.5$   &   $0.5$ & $(5.66 \pm 0.91)\cdot 10^{-3}$  \\
   2  &  $0.7 - 1.0$   &   $1.0$ & $(1.16 \pm 0.15)\cdot 10^{-2}$  \\
   3  &  $1.0 - 1.35$  &   $2.0$ & $(1.34 \pm 0.14)\cdot 10^{-2}$  \\
   4  &  $1.9 - 3.0$   &   $2.0$ & $(4.72 \pm 0.84)\cdot 10^{-3}$  \\
   5  &  $3.0 - 4.5$   &   $2.0$ & $(3.05 \pm 0.80)\cdot 10^{-3}$  \\
\hline
\end{tabular}
\end{table}

Although the systematic trends of the $w(r)$ function are obvious in all the
energy bands, a large scatter of points does not allow for the detailed studies
of the $w(r)$ shape. To quantify a slope of the $w(r)$ in all the energy bands,
we calculate the average values of $w(r)$ in the ranges of $1\arcmin -
2\arcmin$ and $4\arcmin - 7.5\arcmin$. The differences between the $w$
amplitude at these separation bins are given in Table~\ref{delta_w}.  Quoted
uncertainties represent only the statistical scatter of points contributing to
the average values of $w(r)$. Small amplitude of the $w(r)$ results from the
fact that  the total count rates include substantial amount of uniformly
distributed non-cosmic background (especially in the higher energy bands),
and the fluctuations of the XRB are substantially diluted in the raw correlation
function (see Eq.~\ref{ksi_4}).

\section{Sources of the XRB variations \label{models}}

\begin{table*}
\caption{Total observed and model XRB count rates}
\label{countrates}
\centering
\begin{tabular}{clcccccc}
\hline\hline
\noalign{\smallskip}
 $\#$ &     $E$ ~[keV] & \multicolumn{5}{c}
                    {Count Rates ~[cnt\,s$^{-1}$\,pxl$^{-1a}$]} & 
                                                      $\Delta \xi_{\rm PL}$\\
      &                & Total Observed$^b$ & Total Cosmic$^c$  & Power Law$^d$ &
                          WHIM$^e$  & AGN+clusters$^f$ & \\
\hline
\noalign{\smallskip}
   1  &  $0.3 - 0.5$   & $1.13\cdot 10^{-6}$ & $7.9\cdot 10^{-7}$ & $2.3\cdot 10^{-7}$ &
   $(5.2\pm 1.1)\cdot 10^{-8}$ & $4.8\cdot 10^{-7}$ &$0.1419 \pm 0.0228$ \\
   2  &  $0.7 - 1.0$   & $7.27\cdot 10^{-7}$ & $4.7\cdot 10^{-7}$ & $3.7\cdot 10^{-7}$ &
   $(4.1\pm 0.8)\cdot 10^{-8}$ & $4.3\cdot 10^{-7}$ &$0.0445 \pm 0.0058$ \\
   3  &  $1.0 - 1.35$  & $6.73\cdot 10^{-7}$ & $4.2\cdot 10^{-7}$ & $4.2\cdot 10^{-7}$ &
   $(4.8\pm 7.1)\cdot 10^{-9}$ & $4.2\cdot 10^{-7}$ &$0.0337 \pm 0.0035$ \\
   4  &  $1.9 - 3.0$   & $9.81\cdot 10^{-7}$ & $3.8\cdot 10^{-7}$ & $3.8\cdot 10^{-7}$ &
   $ 0 $                       & $3.8\cdot 10^{-7}$ &$0.0310 \pm 0.0055$ \\
   5  &  $3.0 - 4.5$   & $1.04\cdot 10^{-6}$ & $2.9\cdot 10^{-7}$ & $2.9\cdot 10^{-7}$ &
   $ 0 $                       & $2.9\cdot 10^{-7}$ &$0.0383 \pm 0.0101$ \\
\hline
\noalign{\smallskip}
\multicolumn{8}{p{157mm}}{$^a$ pxl = $4\arcsec \times 4\arcsec$;
$^b$ Count rates averaged over all the pointings includeing non-cosmic
     background;
$^c$ Best estimate approximated by the broken power law (see text); 
$^d$ With photon index $\Gamma = -1.42$ and absorption by the cold gas with
$N_{\rm H} = 2.2\cdot 10^{20}$cm$^{-2}$;
$^e$ Count rates actually detected (lower limit for the total WHIM emission);
$^f$ Extragalactic component excluding WHIM.} 
\end{tabular}
\end{table*}

The total count rates recorded in a single pointing are a mixture of the cosmic
signal and various components of the local origin including X-rays produced in
the satellite and the particle background. For each observation a contribution
of the contaminating counts is different and essentially unknown. In the
present investigation the cosmic signal in the individual pointing is not
isolated from the total counts. The average contribution of the non-cosmic
count rates in the whole set of observations is estimated by comparison of the
total count rates in the whole sample with estimates of the cosmic XRB taken
from the literature.

Apart from the small scale fluctuations investigated in the present paper, the
XRB above $1$\,keV is highly isotropic and well approximated by a power law.
Both the normalization and slope have been determined in the past by several
groups using different instruments. In the present analysis we use
parametrization of the XRB by \cite{lumb02} with the photon spectral index
$\Gamma = -1.42$ and the normalization at $1$\,keV of
$9.0$\,ph\,keV$^{-1}$cm$^{-2}$s$^{-1}$sr$^{-1}$.  Below $1$\,keV some fraction
of the XRB is produced by the hot gas in the Galaxy and the WHIM. Galactic
contribution to the sub-keV background exhibits variations over the celestial
sphere, but one should expect that in the arcmin scales this component is
constant. The pointings used in our investigation are scattered over a large
area at high latitudes and it is justified to use the average XRB spectrum
derived by \citet{lumb02}. In the energy ranges of two soft bands
$0.3-0.5$\,keV and $0.7-1$\,keV the cosmic spectrum is adequately
approximated by a power law with the photon index $\Gamma = -2.78$.

The broken power law with these spectral indices is used to calculate the
expected XRB count rates in the EPIC MOS observations in five energy bands.
These count rates are listed in  Table~\ref{countrates} together with the
average total count rates actually observed in the data.

\subsection{Fluctuations above $1$\,keV}

The XRB above $\sim\!\!1$\,keV is generated virtually exclusively by the
extragalactic discrete sources. Most of the flux comes from the whole variety
of AGNs. Some, not well constrained fraction of the XRB is produced by clusters
of galaxies. Neither theoretical considerations, nor observations indicate
that a measurable contribution results from the WHIM emission at these
energies. It is also established that plasma in the Galaxy does not
contribute substantially to XRB above $1$\,keV (e.g. \citealt{galeazzi07},
\citealt{henley07}).
A discrete nature of the XRB at these energies has implications
for the XRB structure. Assuming that the specific population of objects
generates a constant fraction of the XRB in the consecutive energy bands, the
relative amplitude of the XRB fluctuations also should be constant. In the 
following we use this conjecture to analyze the AGNs contribution to the XRB.

Count rates generated by the power law spectrum with $\Gamma = -1.42$ and the
\citet{lumb02} normalization are listed in Table~\ref{countrates}. The low
energy absorption by a cold gas in the Galaxy have been accounted for assuming
the average in the sample hydrogen column density of $2.2\cdot
10^{20}$\,cm$^{-2}$.  In all five energy bands the power law count rates are
substantially lower than the total average count rates recorded in the sample.
In three high energy bands these differences results solely from the non-cosmic
counts, while in two low energy bands also the effect of the plasma emission is
visible. Substituting into Eq.~\ref{ksi_4} the `Total observed' and `Power law'
count rates given in the Table~\ref{countrates} and $\Delta w$ from
Table~\ref{delta_w} we get the expected increments of the ACF, $\Delta \xi_{\rm
PL}$, under the assumption that all the XRB fluctuations are produced solely by
the discrete extragalactic sources generating the power law component. The
model amplitudes of $\Delta \xi_{\rm PL}$ are listed in the last column in
Table~\ref{countrates}.  For three energy bands above $1$\,keV the ACF signal
is consistent with the constant value of $\Delta \xi_{\rm PL} = 0.033$.
However, the model at lower energies requires much larger amplitudes of $\Delta
\xi_{\rm PL}$. It implies that the simple power law model is unable to explain
the total observed amplitude of the XRB fluctuations over the entire energy
range.

\begin{figure}
\resizebox{\hsize}{!}{\includegraphics{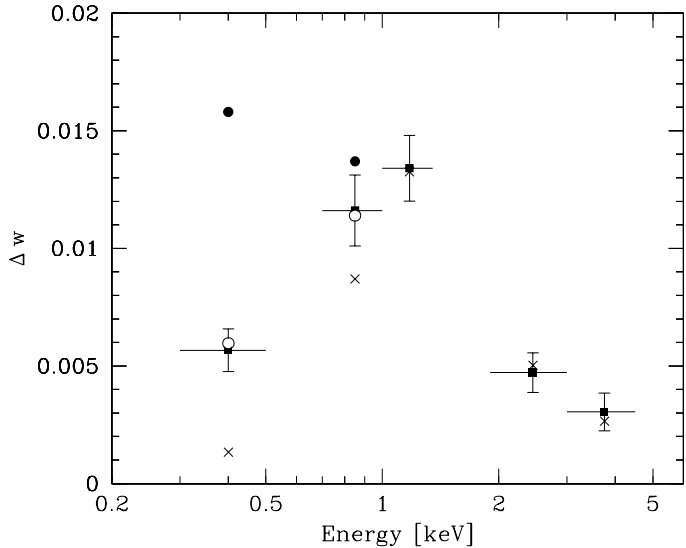}}
\caption{The raw ACF increment $\Delta w$ - points with error bars. Models of
$\Delta w$ distribution assuming $\Delta \xi = 0.033$: crosses - for
fluctuations of power law component of the XRB; dots - for fluctuations of the
whole XRB; open circles - for fluctuations of the XRB component described by
the broken power law (see text for details).}
\label{delta_ksi}
\end{figure}

The relationship between the fluctuations of the raw count rates, $\Delta w$,
and the power law model is visualized in Fig.~\ref{delta_ksi}. Here the model
ACF amplitude, $\Delta \xi$, is used to reproduce the observed fluctuations,
$\Delta w$. In agreement with the above conclusion, the ACF amplitude $\Delta
\xi_{\rm PL} = 0.033$ fits perfectly the $\Delta w$ values in three high energy
bands and fails to reproduce the observed $\Delta w$ signal in two low energy
bands. Higher fluctuation amplitude below $1$\,keV indicates that either the
extragalactic component of the XRB is not adequately represented by a single
power law, or substantial fraction of fluctuations is generated by the WHIM.

\subsection{WHIM contribution to the soft XRB}

Observational constraints on the WHIM contribution to the soft XRB are rather
weak.  The lower limit for the total WHIM emission is defined by the amplitude
of the WHIM signal correlated with the sample of galaxies as measured by
\citet{soltan06}.  In that paper the extended emission in all five energy bands
correlated with galaxies in the local universe was determined. The galaxies
used in the investigation constituted a sample of well-defined statistical
characteristics with surface density of $340$ objects per sq.\,deg.  The
correlated signal corrected for the residual cluster contribution (see the
original paper for details) was used to estimate the mean emission per galaxy
in the sample. This mean flux is used here to calculate the average WHIM
surface brightness which is given in Table~\ref{countrates}. Quoted errors
represent only statistical uncertainties. The WHIM signal in two low energy
bands is detected at better than $4\sigma$ level and constitutes nearly $7$\,\%
of the XRB in the band $0.3-0.5$\,keV and $9$\,\% in $0.7-1.0$\,keV. These
figures represent only the lower limit for the WHIM emission, since basically
unknown additional flux could be potentially generated by the WHIM surrounding
galaxies not represented in the original sample.

The absolute upper limit for the WHIM flux is obtained by minimizing the
contributions due to the galactic plasma and discrete extragalactic sources.
Assuming that below $1$\,keV the average AGN spectrum could still be
approximated by the power law with $\Gamma = -1.42$ and absorption by
a cold gas, the maximum count rate of
the WHIM in the $0.7 - 1.0$\,keV band amounts to $\sim\!1\cdot
10^{-7}$\,cnt\,s$^{-1}$\,pxl$^{-1}$, or $21$\,\% of the XRB (pxl
$= 4\arcsec\times 4\arcsec$). This figure is
obtained assuming negligibly small galactic emission in that band. Assuming
further that the total WHIM emission has a spectral shape similar to the
emission correlated with the galaxies investigated by \citet{soltan06}, we find
that the maximum count rate possibly attributed to the WHIM in the lowest
energy band of $0.3 - 0.5$\,keV amounts to $\sim\!1.3\cdot
10^{-7}$\,cnt$^{-1}$\,pxl$^{-1}$, or $16$\,\% of the XRB.

Although the WHIM emission correlated with galaxies generates a relatively
strong signal in the CCF, the WHIM contribution to the ACF of the integral
background is expected to be small. The available data are insufficient to
determine precisely the ACF amplitude due to the WHIM. Nevertheless, the WHIM
properties as determined by \citet{soltan06} provide some estimates on
the X-ray fluctuation amplitude generated by the intergalactic medium.
The average X-ray surface brightness enhancement around a randomly chosen
galaxy in the sample investigated by \citet{soltan06} does not exceed
$\delta\rho = 3\cdot 10^{-8}$\,cnt\,s$^{-1}$\,pxl$^{-1}$ in the two low energy
bands and is lower in the remaining bands. So, it is below $3$\,\% of the raw
count rate in the $0.3-0.5$\,keV and is around $4$\,\% in the
$0.7-1.0$\,keV band. Since the amplitude of the ACF $w \sim (\delta\rho /
\langle t\rangle)^2$, the WHIM contribution to the increment of $w$ could be
below $0.001$ and $0.002$ in these two bands, respectively. These figures are
smaller than uncertainties of our $\Delta w$ estimates shown in Fig.~\ref{delta_ksi}
and could be neglected in the present considerations. One should note,
however, that allowing a moderate scatter of the halo luminosities, the ACF
signal generated by the WHIM would exceed the above estimates.
Although the low contribution of the WHIM to the fluctuations is favored,
the potential effect of the WHIM ACF uncertainties is accounted for in the
following calculations.

\subsection{Fluctuations below $1$\,keV}

A simple model of the XRB fluctuations generated by a population of sources
with a power law spectrum provides too low amplitude of the fluctuations in the
sub-keV domain. The discrepancy between the observed amplitude $\Delta w$ and
the model implies that either the actual contribution of AGNs to the XRB below
$1$\,keV is higher than the pure power law, or the amplitude of the WHIM
fluctuations has been underestimated.

One should note, however, that the present investigation confirms a well
established fact that the soft XRB cannot be produced entirely by AGNs, but
some fraction of the XRB is generated by a smoothly distributed gas in the
Galaxy.  The total cosmic count rates and $\Delta \xi = 0.033$ would generate
the fluctuations of the raw counts substantially above the observed amplitude
in two low energy bands. This is shown in Fig.~\ref{delta_ksi} with full dots.
Clearly, only a fraction of sub-keV counts is distributed according to $\Delta
\xi = 0.033$, while the remaining counts are distributed more uniformly.

Below we consider two extreme cases of the WHIM contribution to the XRB
fluctuations. If the WHIM emission generates weak fluctuations of the raw
counts, only the discrete extragalactic sources could generate the observed
$\Delta w$ amplitude. In this case, the requirement that the observed
fluctuations below $1$\,keV are reproduced by $\Delta\xi = 0.033$ implies a
softer source spectrum than that at higher energies. Assuming the power law we
get the photon index of $-2.25$. The model which fits the data is shown in
Fig.~\ref{delta_ksi} with open circles and the corresponding count rates are
given in the Table~\ref{countrates}. A conclusion that the fluctuating XRB
component exhibits soft excess is in agreement with the steepening of the AGN
spectra below $1$\,keV observed in individual X-ray bright objects (see for
example \citealt{elvis94}). Similar effect is visible also in the
population of weaker sources which generate $\sim 30$\,\% of the XRB at
$2$\,keV \citep{mateos05}.  It is possible, however, that this apparent
coincidence is fortuitous, since the XRB spectrum is significantly harder than
the average spectrum  of the AGNs investigated by those authors. Thus, the
remaining $70$\,\% of the XRB is produced by still weaker sources which do not
necessarily exhibit the soft excess.

If the integrated AGN contribution to the XRB below $1$\,keV does not exceed
the power law extrapolated from the higher energies, the fluctuation amplitude
in the sub-keV domain could be generated by poor clusters or groups of galaxies
with masses below $\sim 2\cdot 10^{13}$\,M$_{\odot}$. Since the gas temperature
in these objects is typically lower than $1$\,keV \citep{horner99}, their
extended emission contributes to the fluctuations at scales of a few arcmin
merely in the soft energy bands. Unfortunately, the present estimates of the ACF
shape are inadequate to distinguish between different fluctuation mechanisms
and we are unable isolate the cluster contribution.

The soft excess observed in the extragalactic component reduces
estimates of the galactic signal. The difference between the total cosmic
signal and the extragalactic component provides an estimate of the mean
galactic contribution to the XRB. The extragalactic broken power law together
with the observed WHIM flux generate all the XRB in the $0.7-1.0$\,keV band and
$67$\,\% of the XRB in the $0.3-0.5$\,keV band. Thus, the galactic component
visible only in the softest band, is substantially softer of what is
usually assumed.

The relationship between various components is not significantly changed if
the average spectrum of extragalactic discrete sources over the entire energy
range is power law with $\Gamma = -1.42$ and the ``missing'' $\Delta w$ amplitude
below $1$\,keV is generated by the WHIM. In this case the galactic plasma
would contribute up to $65$\,\% in the $0.3-0.5$\,keV band and to $13$\,\% in the
$0.7-1.0$\,keV band. Although this contribution is substantially higher than
in the broken power law case, the galactic spectrum is still very soft.

\section{Discussion and conclusions \label{conclusions}}

Absolute fluxes of the major components of the soft XRB are investigated using
the information extracted from the correlation functions. The variations of the
correlation amplitude with photon energy have allowed us to separate the
fluctuating extragalactic component from the smooth (although also variable at
large scales) galactic emission. The total fluctuations above $1$\,keV are
adequately described by a single ACF. This conclusion is in agreement with a
well-established fact that at high energies the XRB is generated entirely by
the extragalactic discrete sources (AGNs and clusters).  One should note that
the present estimate of the increment, $\Delta \xi = 0.033$, is in good
agreement with the \citet{soltan01} measurement of the XRB fluctuations based
on the {\it ROSAT} pointings \citep{soltan07}.

In calculations below $1$\,keV we apply the fluctuation analysis without any a
priori assumption on the spectral shape of both components, while in the
`canonical' approach to the question of the galactic contribution, the power
law spectrum of the extragalactic component is usually assumed (cf
\citealt{galeazzi07, henley07}). This is justified at energies above $1$\,keV,
but in the sub-keV region it is likely that the power law approximation fails
to represent the average spectrum of AGNs.

The WHIM contribution to the total XRB flux is distinctly smaller than the AGN
component. But the soft WHIM spectrum substantially affects the XRB budget
below $1$\,keV. Because the galactic emission is not easily distinguished from
the WHIM signal, the soft section of the XRB was frequently attributed to the
galactic plasma ignoring the genuine contribution by the WHIM.  The lower limit
for the WHIM flux is determined by the cross-correlation of the XRB surface
brightness with the local galaxy distribution \citep{soltan06}. The total
WHIM contribution to the XRB is assessed at $7-16$\,\% in $0.3-0.5$\,keV band
and $9-21$\,\% in $0.7-1.0$\,keV.  In the sub-keV region both galactic hot
plasma and the WHIM contribute to the XRB. Nevertheless, in the band of
$0.7-1.0$\,keV, the AGNs generate still at least $79$\,\% of the total XRB
flux. The extragalactic discrete source contribution drops significantly below
$0.5$\,keV, but even in the softest band it is likely that AGNs make up to
$60$\,\% of the background.

It is estimated that the galactic emission averaged over many pointings
scattered in two galactic hemispheres generates $33$\,\% of the total cosmic
signal in the softest band (although a higher value is not ruled out in the
present investigation).  The galactic contribution drops sharply with the
energy. In the $0.7-1.0$\,keV band it amounts to at most $13$\,\% of the XRB,
and the data are consistent with no detectable galactic signal. It appears that
the mean temperature of the emitting plasma does not exceed significantly
$10^6$\,K. It roughly corresponds to the estimates of the temperature in the
Local Bubble, but is inconsistent with the hot galactic halo model. According
to the standard approach, the entire excess of the soft XRB above the power law
extrapolation is attributed to the Galaxy. Detailed studies by several groups
have measured the Local Bubble emission and estimated its temperature at about
$10^6$\,K.  Thermal emission of such plasma, however, cannot explain the signal
between $\sim0.6$ and $1$\,keV. To account for the excess flux in this energy
range, it is postulated that the Galaxy is surrounded by the halo of hot plasma
with the temperatures close to $3\cdot 10^6$\,K. Our analysis indicates that
the amplitude of the hot halo emission is substantially smaller than it is
usually assumed. The discrepancy between the present results and earlier
estimates results directly from the analysis of the WHIM emission
and the different estimates of the extragalactic component.

\begin{acknowledgements}
I thank all the people involved in the XMM-Newton project for making the XMM
Science Archive and Standard Analysis System such user-friendly environment.
This work has been partially supported by the Polish KBN grant 1~P03D~003~27.
\end{acknowledgements}

\end{document}